\documentclass[journal]{IEEEtran}
\ifCLASSINFOpdf
\else
\fi
\usepackage{amsmath}
\usepackage{pgfplots}
\usepackage{amsfonts}
\pgfplotsset{compat=1.16}
\usepackage{tikz}
\usepackage{multicol}

\begin{document}

\title{Theory of Periodically Time-Variant Linear Systems}

\author{\IEEEauthorblockN{{Juan I. Bonetti}\IEEEauthorrefmark{1},
    {Agustín Galetto\IEEEauthorrefmark{1}, and Mario R. Hueda}\IEEEauthorrefmark{2}}\\
  \IEEEauthorblockA {
    \IEEEauthorrefmark{1} Fundaci\'on Fulgor - Romagosa 518 - C\'ordoba (5000) - Argentina \\
    \IEEEauthorrefmark{2} Laboratorio de Comunicaciones Digitales - Universidad Nacional de  C\'ordoba\\
    Av. V\'elez Sarsfield 1611 - C\'ordoba (X5016GCA) - Argentina\\
    Email: juan.bonetti@ib.edu.ar}}


\maketitle

\begin{abstract}
  In this work we provide a mathematical framework to describe the periodically time variant (PTV) linear systems. We study their frequency-domain features to estimate the output bandwidth, a necessary value to obtain a suitable digital representation of such systems. In addition, we derive several interesting properties enabling useful equivalences to represent, simulate and compensate PTVs.
\end{abstract}

\IEEEpeerreviewmaketitle

\section{Definition}
\begin{figure}
    \centering
    \includegraphics[scale=0.5]{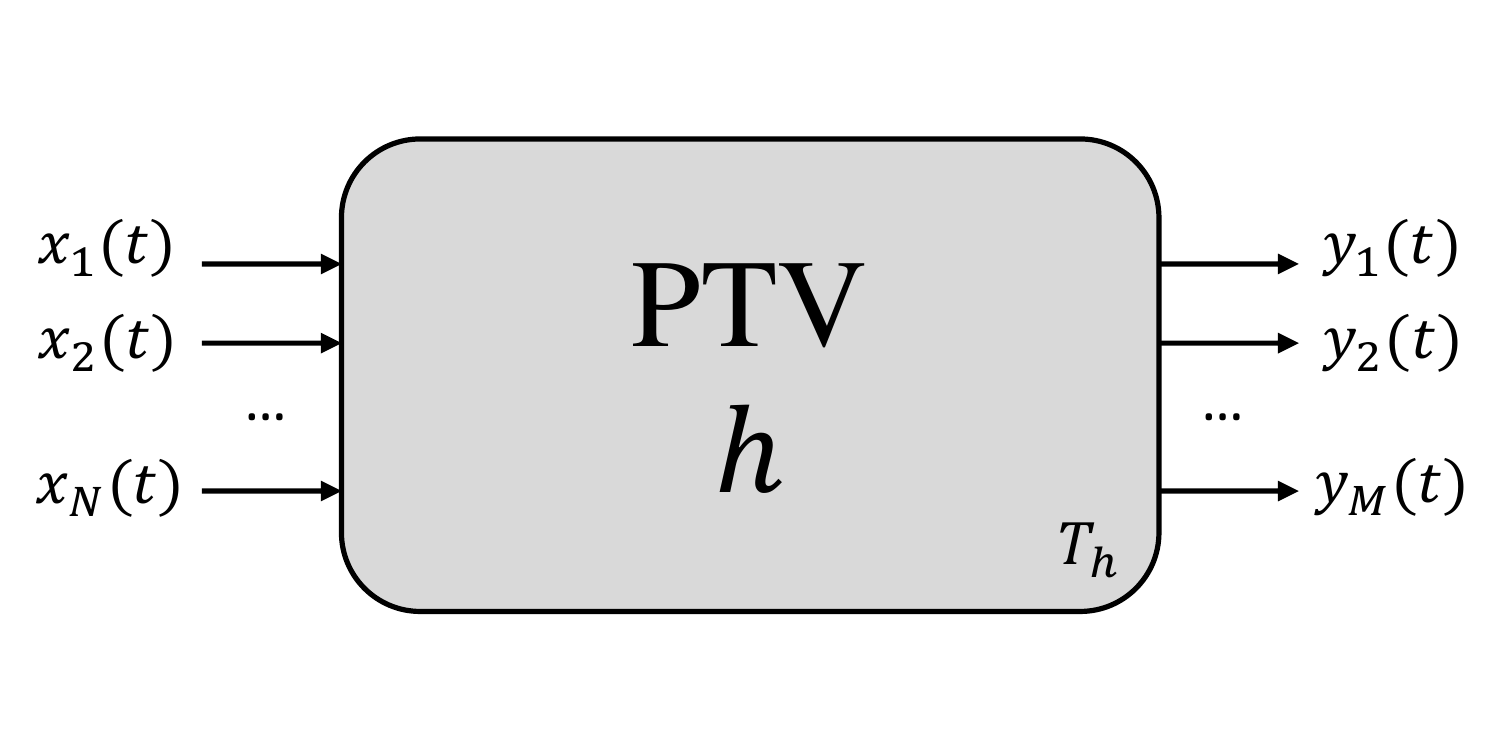}
   \caption{Representation of a generic PTV linear system, relating $N$ continuous-time inputs $x_j(t)$ with $M$ continuous-time outputs $y_i(t)$. While $h$ is the name of the PTV, $T_h$ stands for its period.}
\label{fig01}
\end{figure}

A time-variant (TV) linear system is defined by an impulse response that depends on time. In general, a TV linear system of $N$ inputs and $M$ outputs can be written as~\cite{claasen1982on,middleton1988adaptive}
\begin{equation}
    y_i(t) = \sum_{j=1}^{N} \int h_{ij}(t,\tau)x_j(t-\tau)\,d\tau \quad \forall i\in\{1,2,...,M\},
    \label{def1}
\end{equation}
being $x_j(t)$ and $y_i(t)$ the continuous-time system inputs and outputs, respectively, and $h_{ij}(t,\tau)$ the impulse responses. A \textit{periodically time-variant} (PTV) linear system is a TV system whose impulse responses present a periodic behavior in the time variable, \textit{i.e.,}
\begin{equation}
    h_{ij}(t+T_h,\tau) = h_{ij}(t,\tau).
    \label{def2}
\end{equation}
being $T_h$ the PTV period. Figure~\ref{fig01} shows the schematic representation of the PTV $h$, described by Eqs.~\ref{def1}~and~\ref{def2}. We introduce the variable \textit{temporal phase} $z_h$, defined as $z_h(t) = \mathrm{mod}(t,T_h)$, being mod the modulo operation. The temporal phase allows for the definition of the PTV from simplified impulse responses,
\begin{equation}
    y_i(t) = \sum_{j=1}^{N} \int h_{ij}(z_h(t),\tau)x_j(t-\tau)\,d\tau \quad \forall i\in\{1,2,...,M\},
    \label{def3}
\end{equation}
as the first argument of $h_{ij}$ is restricted to values between $0$ and $T_h$.

\begin{figure}
    \centering
    \includegraphics[scale=0.35]{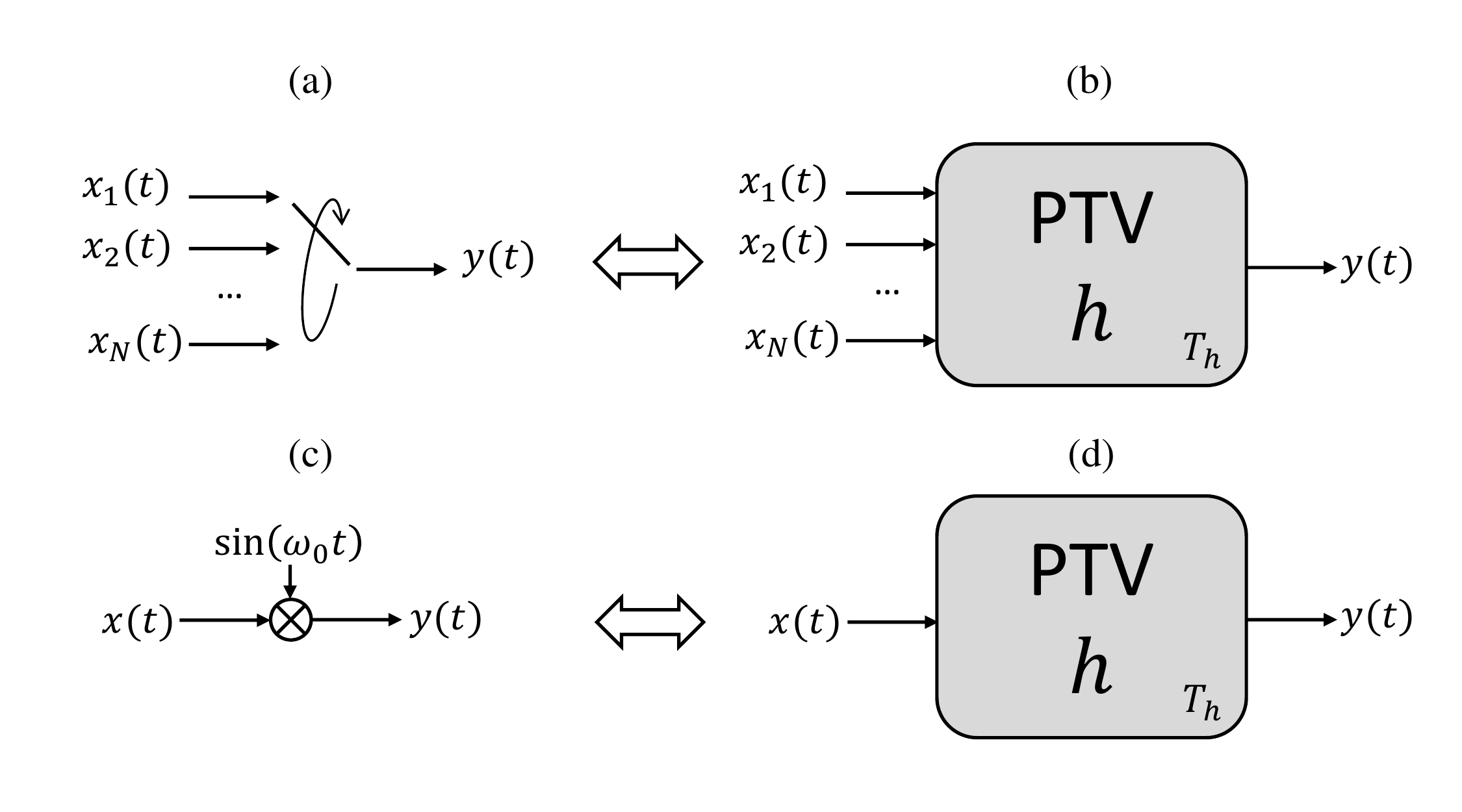}
   \caption{Examples of common PTVs: a) cyclical multiplexer $N:1$. b) equivalent $N:1$ PTV model. c) local oscillator multiplier. d) equivalent PTV model.}
\label{fig02}
\end{figure}

A clear example of PTV system is the ideal cyclical multiplexer $N:1$, shown in Fig.~\ref{fig02}(a), a device that periodically alternates its single output between its $N$ inputs. As shown in Fig.~\ref{fig02}(b), it can be modeled as a single-output PTV, of period $T_h$, given by
\begin{equation}
    y(t) = \sum_{j=1}^{N} \int h_{j}(z_h(t),\tau)x_j(t-\tau)\,d\tau,
\end{equation}
where the impulse responses are
\begin{equation}
    h_{j}(z,\tau) = \left\lbrace \begin{array}{l} \delta(\tau),\\0, \end{array}\right. \quad \begin{array}{c} (j-1)T_h/N\leq z < jT_h/N\\ \mathrm{otherwise,}\end{array}
\end{equation}
being $\delta(.)$ the Dirac delta function. Another common example is the multiplier with a local oscillator input, displayed in Fig.~\ref{fig02}(c). Although the output can be easily written as $y(t) = x(t)\mathrm{sin}(\omega_0 t)$, we can take it to the PTV form, as shown in Fig.~\ref{fig02}(d), by writing
\begin{equation}
    y(t) = \int h(z_h(t),\tau)x(t-\tau)\,d\tau,
    \label{localosc1}
\end{equation}
with
\begin{equation}
    h(z,\tau) = \delta(\tau)\mathrm{sin}(\omega_0 z)
    \label{localosc2}
\end{equation}
and $T_h = 2\pi/\omega_0$.

\section{Combination of PTVs}
In this section we study the interaction between PTVs and time-invariant linear systems.
\subsection{Parallel PTVs}
\begin{figure}
    \centering
    \includegraphics[scale=0.3]{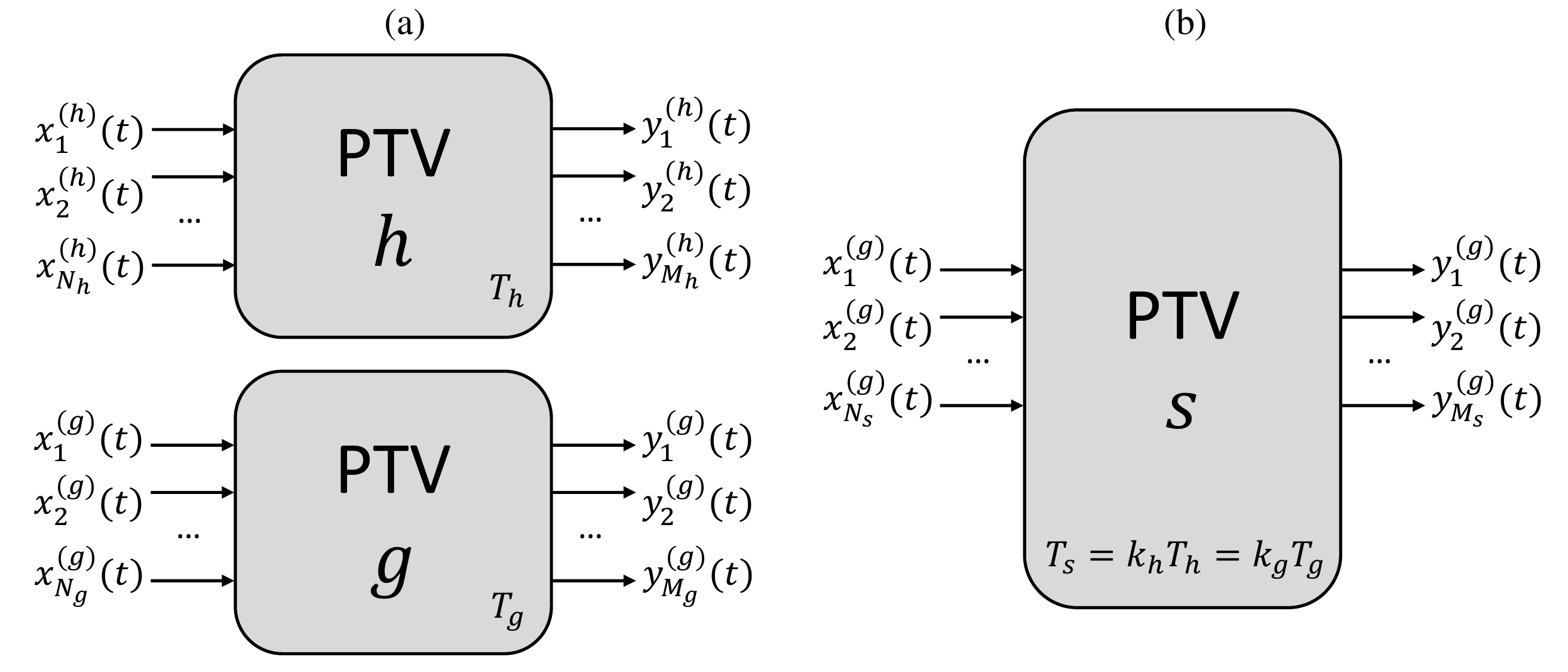}
   \caption{Parallel configuration of PTVs: a) the systems $h$ and $g$ are PTVs of period $T_h$ and $T_g$, respectively.  b) equivalent PTV model $s$ of period $T_s$, the least common multiplier of $T_h$ and $T_g$.}
\label{fig03}
\end{figure}

We consider the two PTV systems $h$ and $g$, shown in Fig.~\ref{fig03}(a), described by the set of equations
\begin{equation}
    y^{(h)}_i(t) = \sum_{j=1}^{N_h} \int h_{ij}(z_h(t),\tau)x^{(h)}_j(t-\tau)\,d\tau,
\end{equation}
with $i\in\{1,2,...,M_h\}$, and
\begin{equation}
    y^{(g)}_i(t) = \sum_{j=1}^{N_g} \int g_{ij}(z_g(t),\tau)x^{(g)}_j(t-\tau)\,d\tau,
\end{equation}
with $i\in\{1,2,...,M_g\}$, respectively. We define a new time-variant system, $s$, whose inputs/outputs are given by
\begin{equation}
    x_j(t) = \left\lbrace \begin{array}{l} x^{(h)}_j(t),\\x^{(g)}_{j-N_h},\end{array} \right. \quad \begin{array}{c} j\in\{1,2,..,N_h\}\\j\in\{N_h+1,N_h+2,..,N_h+N_g\}, \end{array}
\end{equation}
and
\begin{equation}
    y_i(t) = \left\lbrace \begin{array}{l} y^{(h)}_i(t),\\y^{(g)}_{i-M_h},\end{array} \right. \quad \begin{array}{c} i\in\{1,2,..,M_h\}\\i\in\{M_h+1,M_h+2,..,M_h+M_g\}. \end{array}
\end{equation}
The system $s$ is then described by the linear relationship
\begin{equation}
    y_i(t) = \sum_{j=1}^{N_s} \int s_{ij}(t,\tau)x_j(t-\tau)\,d\tau \quad \forall i\in\{1,2,...,M_s\},
\end{equation}
where $N_s = N_h + N_g$, $M_s = M_h + M_g$, and
\begin{equation}
    s_{ij}(t,\tau) = \left\lbrace \begin{array}{l} h_{ij}(z_h(t),\tau),\\ \qquad i\in\{1,2,...,M_h\}\land j\in\{1,2,...,N_h\}\\ g_{(i-M_h)(j-N_h)}(z_g(t),\tau),\\ \qquad i\in\{M_h+1,M_h+2,...,M_s\}\land\\ \qquad j\in\{N_h+1,N_h+2,...,N_s\}\\ 0, \quad \mathrm{otherwise.} \end{array} \right.
    \label{sdef}
\end{equation}
If two integers, $k_h$ and $k_g$, can be found to satisfy $k_hT_h = k_gT_g$, the system $s$ is also a PTV. The period of $s$ is given by
\begin{equation}
    T_s = k_hT_h = k_gT_g,
    \label{lcm}
\end{equation}
being $k_h$ and $k_g$ the minimum integers satisfying the equality. In other words, if $T_s$ exists, can be obtained as the least common multiple (lcm) of the periods $T_h$ and $T_g$. Note that $T_s$ exists only if $T_h/T_g$ is a rational number. The PTV behavior of $s$ can be easily proven with Eq.~\ref{sdef}, obtaining
\begin{equation}
     s_{ij}(t+T_s,\tau) = s_{ij}(t,\tau) = s_{ij}(z_s(t),\tau).
     \label{s_ptv}
\end{equation}
Figure~\ref{fig03}(b) shows the equivalent PTV system $s$ resulting from the parallel topology of $h$ and $g$. 

\subsection{Series PTVs}
\begin{figure}
    \centering
    \includegraphics[scale=0.31]{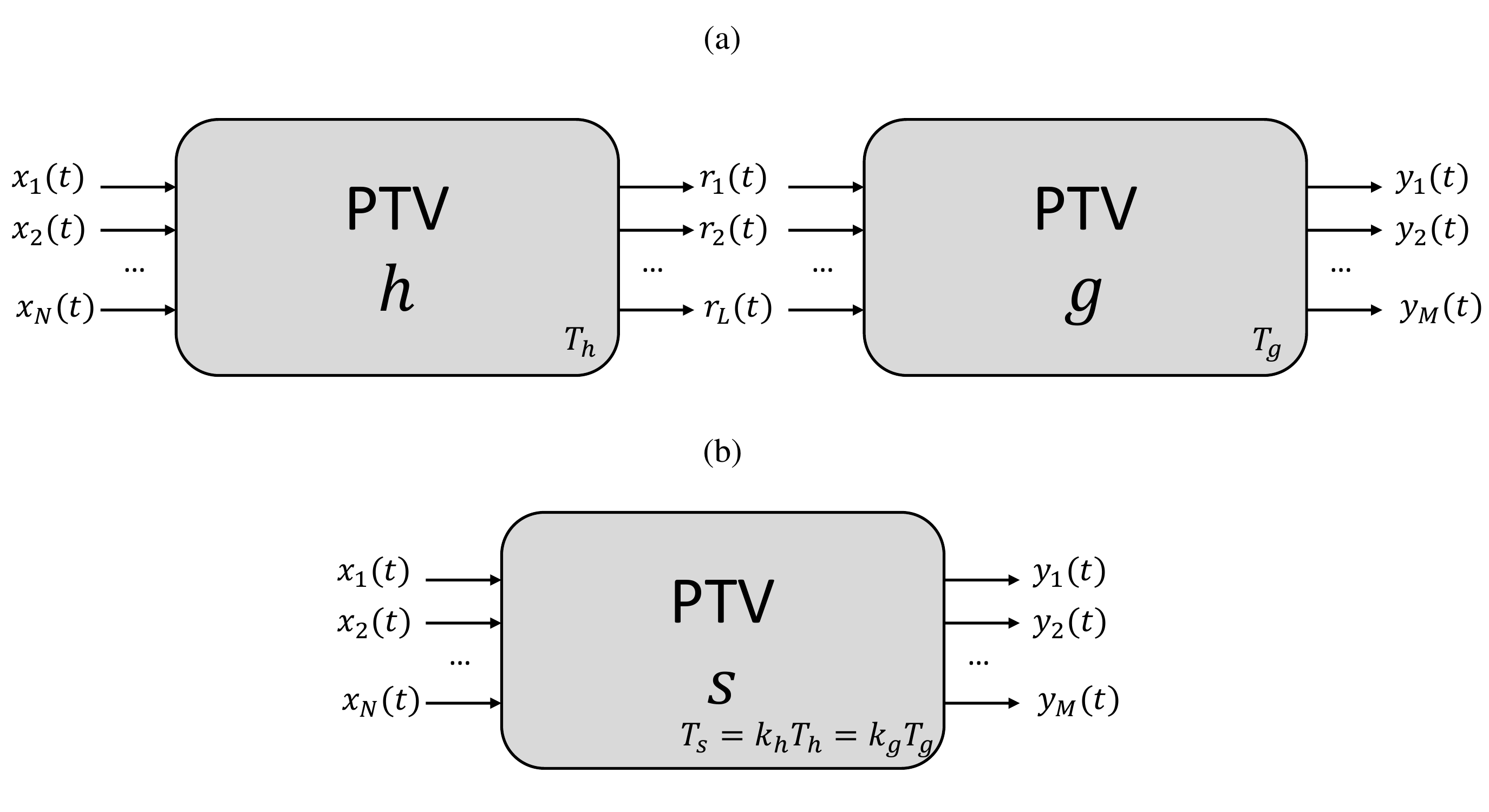}
   \caption{Series configuration of PTVs: a) the systems $h$ and $g$ are PTVs of period $T_h$ and $T_g$, respectively.  b) equivalent PTV model $s$ of period $T_s$, the least common multiplier of $T_h$ and $T_g$.}
\label{fig04}
\end{figure}

In the series configuration of the PTVs $h$ and $g$, shown in Fig.~\ref{fig04}(a), the $L$ outputs of system $h$ are the inputs of the system $g$. By combining the input-output equations of both systems,
\begin{equation}
    r_l(t) = \sum_{j=1}^{N} \int h_{ij}(z_h(t),\tau)x_j(t-\tau)\,d\tau \quad \forall l\in\{1,2,...,L\}
\end{equation}
and
\begin{equation}
    y_i(t) = \sum_{l=1}^{L} \int g_{il}(z_g(t),\tau)r_l(t-\tau)\,d\tau \quad \forall i\in\{1,2,...,M\},
\end{equation}
we obtain an equivalent TV linear system $s$, given by
\begin{equation}
    y_i(t) = \sum_{j=1}^{N} \int s_{ij}(t,\tau)x_j(t-\tau)\,d\tau \quad \forall i\in\{1,2,...,M\},
\end{equation}
where
\begin{multline}
    s_{ij}(t,\tau) =\\ \sum_{l=1}^L \int g_{il}(\mathrm{mod}(t,T_g),\mu)h_{lj}(\mathrm{mod}(t-\mu,T_h),\tau-\mu) \,d\mu.
\end{multline}
As in the case of the parallel configuration, if the lcm of both periods can be found, $s$ is proven to be a PTV system satisfying Eqs.~\ref{lcm} and~\ref{s_ptv}. Figure~\ref{fig04}(b) displays the equivalent PTV system of the series PTVs.

\subsection{Combination with time-invariant linear systems}
A time-invariant linear system can be expressed as a PTV with an arbitrary period. For instance, the linear system $L$, described by
\begin{equation}
    y_i(t) = \sum_{j=1}^{N} \int L_{ij}(\tau)x_j(t-\tau)\,d\tau \quad \forall i\in\{1,2,...,M\},
\end{equation}
can be also defined as the PTV system $h$, given by Eq.~\ref{def3}, where
\begin{equation}
    h_{ij}(z_h(t),\tau) = L_{ij}(\tau).
    \label{tvtoptv}
\end{equation}
As $h$ does not depend on $z_h(t)$, the period $T_h$ can be arbitrarily set. Consequently, combination of time-invariant linear systems with PTVs can be reduced to a unique PTV system by following the rules of parallel and series configuration introduced before.

\begin{figure}
    \centering
    \includegraphics[scale=0.28]{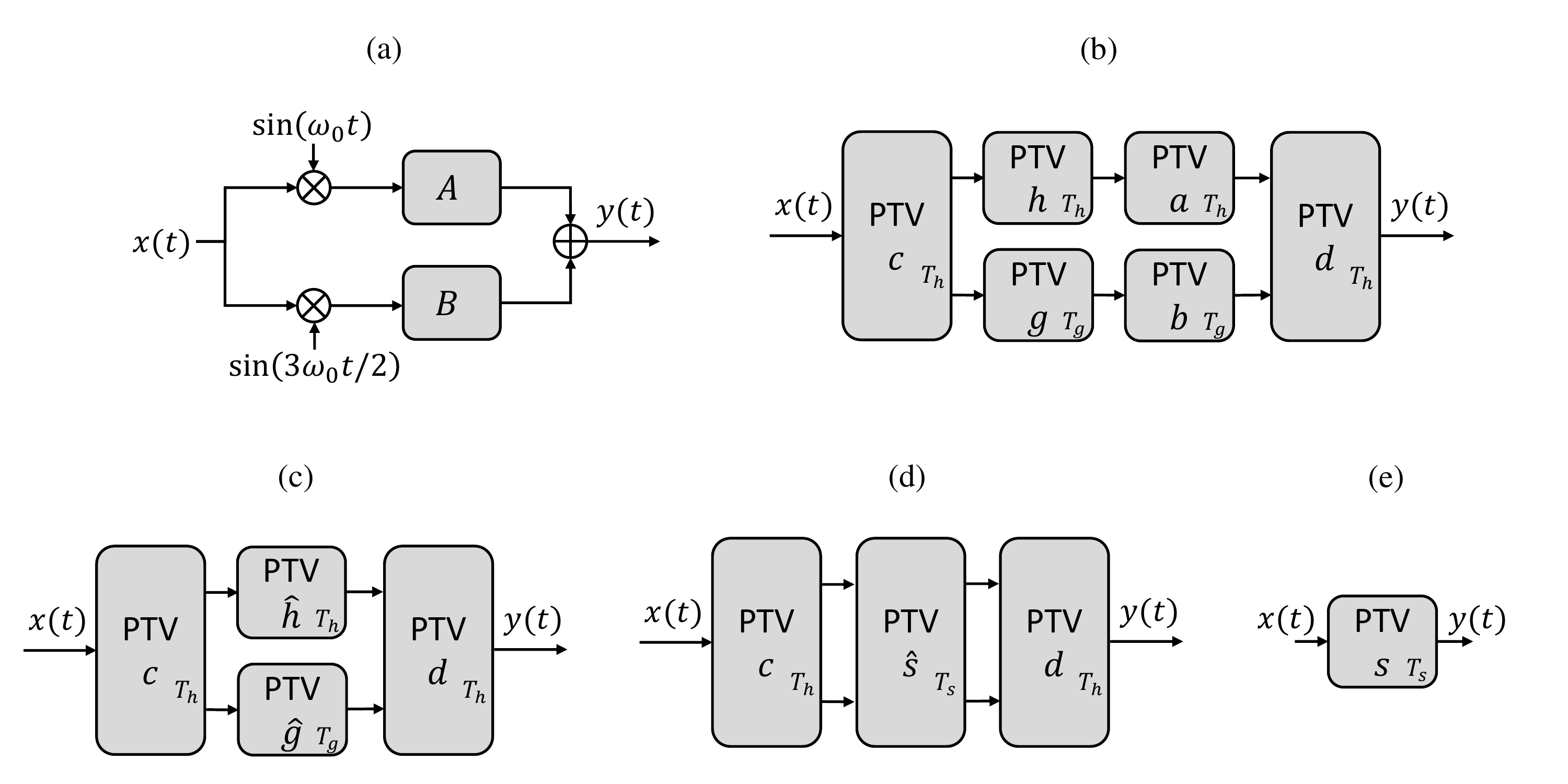}
   \caption{Example of PTV reduction: a) circuit combining PTVs (local oscillator multipliers) with time-invariant linear systems ($A$ and $B$).  b) each component of the circuit is expressed in its PTV form. c) the series PTV $h$-$a$ and $g$-$b$ are reduced to the single PTV form. d) parallel reduction of the system $\hat{h}||\hat{g}$. e) final equivalent PTV system of the circuit.}
   \label{fig05}
\end{figure}

As a simple example, we study the system shown in Fig.~\ref{fig05}(a): a linear combination of two local oscillator multiplier lines. The blocks $A$ and $B$ represent time-invariant linear systems. In Fig.~\ref{fig05}(b) we show the representation of all the circuit components as PTV systems. The local oscillator multipliers are converted to the systems $a$ and $b$ by following Eqs.~\ref{localosc1} and~\ref{localosc2}, and their periods are defined as $T_h = 2\pi/\omega_0$ and $T_g = 4\pi/3\omega_0$, respectively. Systems $A$ and $B$ are regarded as the PTV systems $a$ and $b$ by using Eq.~\ref{tvtoptv}. Their periods are conveniently set to $T_h$ and $T_g$, respectively. Also, the split and sum points are regarded as 1:2 and 2:1 PTVs, respectively, both with period $T_h$. In the next step, shown in Fig.~\ref{fig05}(c), we reduce the series PTVs $h$($g$) and $a$($b$) to the single PTVs $\hat{h}$($\hat{g}$). Then, as shown in Fig.~\ref{fig05}(d), the parallel configuration of $\hat{h}$ and $\hat{g}$ is reduced to the PTV $\hat{s}$. The period $T_s$ can be easily calculated by expressing the period ratio as a fraction:
\begin{equation}
    \frac{T_h}{T_g} = \frac{3}{2} \quad \Leftrightarrow \quad 2 T_h = 3 T_g.
    \label{simplefraction}
\end{equation}
By comparing Eq.~\ref{simplefraction} with Eq.~\ref{lcm}, we obtain $T_s = 4\pi/\omega_0$. Finally, in Fig.~\ref{fig05}(e), we reduce the serie of $c$-$\hat{s}$-$d$ in the 1:1 PTV $s$, whose period can be easily proven to be $T_s$.

\section{Output Bandwidth}
Unlike the time-invariant linear systems, the output bandwidth of a PTV is not necessarily equal to the input bandwidth. A clear example is provided by the case shown in Fig.~\ref{fig02}(c), where the local oscillator multiplier increases the signal bandwidth due to the frequency translation process. In this section we derive a simple formula to calculate the output bandwidth.

At the first place, we note that the frequency-domain representation of the PTV described by the impulse responses $h_{ij}(z,\tau)$ is given by the two-dimensional functions
\begin{equation}
    \tilde{h}_{ij}(k,f) = \int_0^{T_h}\int_{-\infty}^{\infty} h_{ij}(z,\tau) e^{-\text{j}2\pi\left(kz/T_h+f\tau\right)}\,d\tau\,dz,
    \label{fourier}
\end{equation}
where $k\in\mathbb{Z}$ and $f\in\mathbb{R}$. This definition represents an hybrid transformation combining the Fourier transform on $\tau$ with the Fourier series on $z$, due to the periodic behavior of $h_{ij}$ on the last variable. The inverse of Eq.~\ref{fourier} leads to the definition of two bandwidths for the PTV $h$: the \textit{variation bandwidth} $A_h$, corresponding to the discrete variable $k$, and the \textit{linear bandwidth} $B_h$, corresponding to the continuous variable $f$, as the minimum values satisfying
\begin{equation}
    h_{ij}(z,\tau) = \sum_{k=-A_h}^{A_h} \int_{-B_h}^{B_h} \tilde{h}_{ij}(k,f) e^{\text{j}2\pi\left(kz/T_h+f\tau\right)}\,d\tau
\end{equation}
$\forall i,j$. While the linear bandwidth has a simple interpretation as the bandwidth of time-invariant linear systems, the variation bandwidth is a particular property of the PTVs, associated to the maximum variation speed of the impulse-response with respect to the temporal variable. 

By using the definition of Eq.~\ref{fourier} and the inverse Fourier transform of $x_{i}$,
\begin{equation}
    x_{i}(t) = \int_{-B_x}^{B_x} \tilde{x}_{i}(f)e^{\text{j}2\pi f t} \,df,
    \label{invFourier}
\end{equation}
where $\tilde{x}_{i}$ and $B_x$ are the Fourier transform and the bandwidth of $x_{i}$, respectively, in Eq.~\ref{def1} we obtain
\begin{multline}
    y_i(t) = \sum_{j=1}^{N}\sum_{k=-A_h}^{A_h}\int_{-B_x}^{B_x}\int_{-B_h}^{B_h} \int \\ \tilde{h}_{ij}(k,f)\tilde{x}_j(f')e^{\text{j}2\pi\left(kt/T_h + f\tau +f'(t-\tau)\right)}\,d\tau\,df\,df'.
\end{multline}
By making the change of variable $f'=\mu-k/T_h$ we have
\begin{equation}
    y_i(t) = \sum_{k=-A_h}^{A_h}\int_{-B_x+k/T_h}^{B_x+k/T_h}\tilde{y}_{i}(k,\mu)
    e^{\text{j}2\pi\mu t}\,d\mu,
    \label{yfreq}
\end{equation}
where
\begin{multline}
    \tilde{y}_{i}(k,\mu) = \\ \sum_{j=1}^{N} \int_{-B_h}^{B_h} \int \tilde{h}_{ij}(k,f)\tilde{x}_j(\mu-k/T_h)e^{\text{j}2\pi\left(f-\mu+k/T_h\right)\tau}\,d\tau\,df.
\end{multline}
Although Eq.~\ref{yfreq} is not an usual inverse Fourier transform, like Eq.~\ref{invFourier}, it allows for the calculation of the output bandwidth $B_y$, as the maximum-frequency component of $y_i(t)$ is clearly
\begin{equation}
    B_y = B_x + \frac{A_h}{T_h}.
\end{equation}

\section{Discrete-time representation of PTVs}
By knowing the output bandwidth of a PTV, we are able to perform a discrete-time representation of the system. We have to choose a sampling period $T_\mathrm{s}$ satisfying the Nyquist condition, \textit{i.e.}
\begin{equation}
    T_{\mathrm{s}} \leq \frac{1}{2B_y},
\end{equation}
and then to define the discrete-time signals as
\begin{equation}
    a[n] = a(nT_\mathrm{s}),
    \label{sampling}
\end{equation}
where $n\in\mathbb{Z}$ and $a$ stands for any input-output signal. An useful operation is the inverse of the sampling process of Eq.~\ref{sampling}, given by 
\begin{equation}
    a(t) = \sum_{n=-\infty}^{\infty} a[n] \mathrm{sinc}\left(\frac{t}{T_\mathrm{s}} - n\right),
    \label{invsampling}
\end{equation}
where $\mathrm{sinc}(x)=\mathrm{sin}(\pi x)/(\pi x)$ $\forall x\neq 0$ and $\mathrm{sinc}(0) = 1$.

\begin{figure}
    \centering
    \includegraphics[scale=0.5]{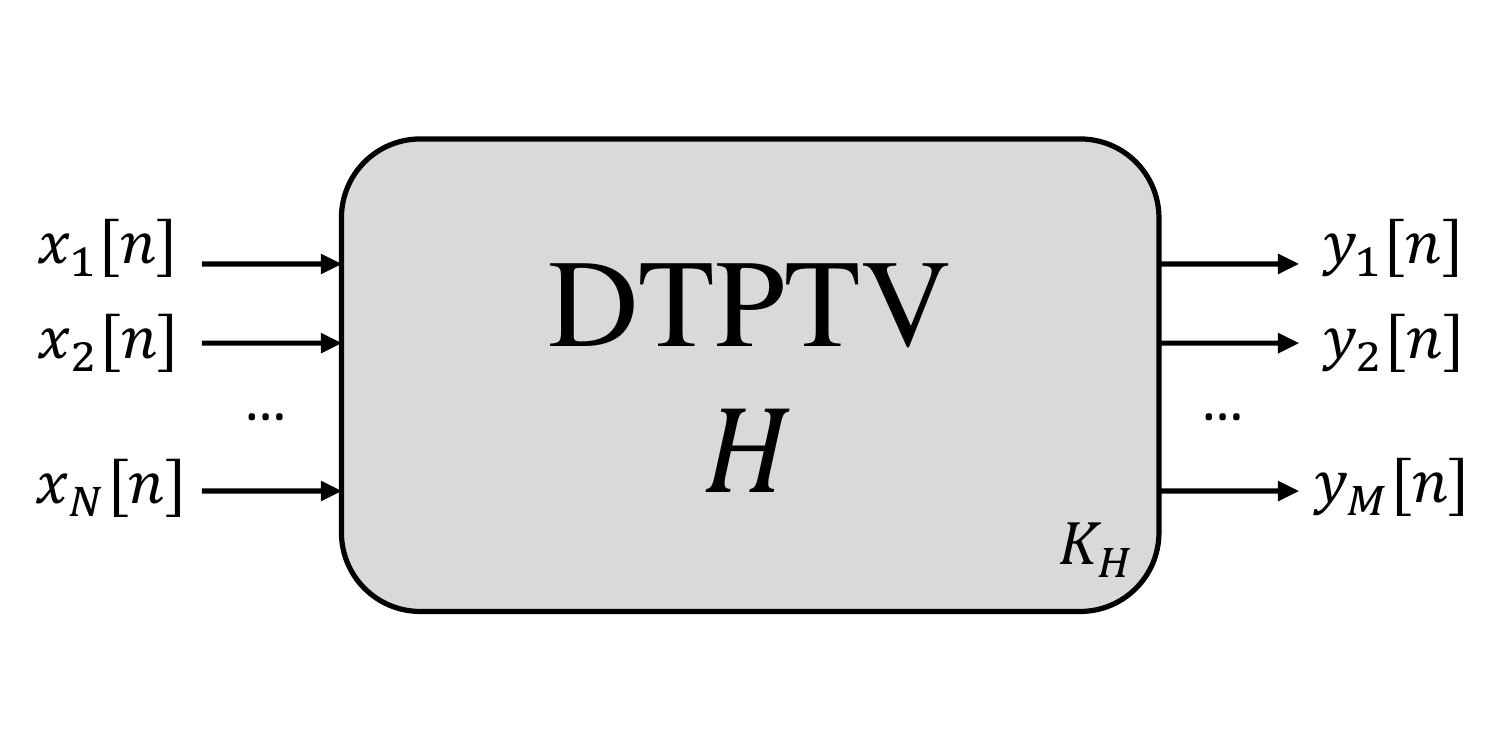}
   \caption{Representation of a generic DTPTV linear system, relating $N$ discrete-time inputs $x_j[n]$ with $M$ discrete-time outputs $y_i[n]$. While $H$ is the name of the DTPTV, $K_H$ stands for its discrete period.}
\label{fig06}
\end{figure}

By using Eqs.~\ref{sampling} and~\ref{invsampling} in the definition of PTV (Eq.~\ref{def3}), we obtain
\begin{equation}
    y_i[n] = \sum_{j=1}^{N} \sum_{m=-\infty}^{\infty} H_{ij}[n,m]x_j[n-m],
    \label{DTTV}
\end{equation}
where 
\begin{equation}
    H_{ij}[n,m] = \int h_{ij}(\mathrm{mod}(nT_\mathrm{s},T_h),\tau) \mathrm{sinc}\left(m-\frac{\tau}{T_\mathrm{s}}\right) \,d\tau.
    \label{Hdisc}
\end{equation}
Equation~\ref{DTTV} is the definition of a discrete-time TV system, as the impulse responses $H_{ij}$ do not only depend on the input sampling index $m$ but also of the output sampling index $n$. In addition, if the sampling period is set to be a divisor of the PTV period, \textit{i.e.}
\begin{equation}
    T_h = K_H T_\mathrm{s} \quad K_H\in\mathbb{Z},
\end{equation}
Eq.~\ref{DTTV} becomes the definition of a \textit{discrete-time periodically time-variant} (DTPTV) linear system, that reads
\begin{equation}
    y_i[n] = \sum_{j=1}^{N} \sum_{m=-\infty}^{\infty} H_{ij}[z_H[n],m]x_j[n-m],
    \label{DTPTV}
\end{equation}
with $z_H[n] = \mathrm{mod}(n,K_H)$ and being $K_H$ the discrete period of the system $H$. Figure~\ref{fig06} shows the schematic representation of a DTPTV system. Equation~\ref{DTPTV} allows for the numerical simulation of PTV system and enables the demonstration of two interesting properties, as shown in next section.

\section{Inverse of PTVs}
We use the discrete-time representation to prove that that the inverse of a PTV linear system, if it exists, is another PTV of the same dimension.

\subsection{SISO PTV}
\begin{figure}
    \centering
    \includegraphics[scale=0.275]{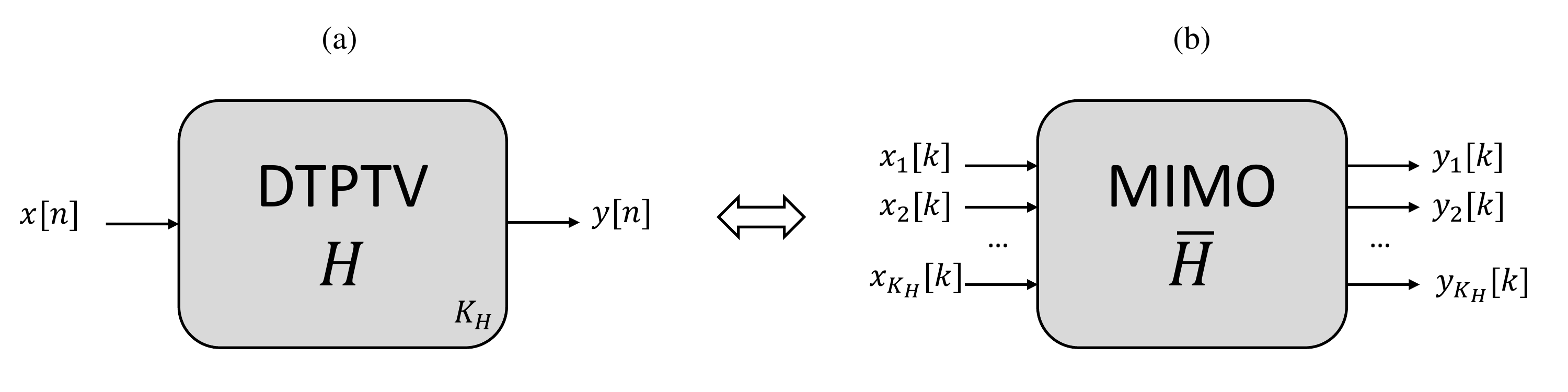}
   \caption{Equivalence between a SISO PTV and a time-invariant MIMO linear system. a) discrete-time of a SISO PTV. b) equivalent MIMO $K_H \times K_H$, obtained by considering vector signals.}
\label{fig07}
\end{figure}

The \textit{single-input single-output PTV} (SISO PTV), shown in Fig.~\ref{fig07}(a), can be expressed in its discrete-time form as
\begin{multline}
    y[n] = \sum_{m=-\infty}^{\infty} H[z_H[n],m]x[n-m] =\\ \sum_{m=-\infty}^{\infty} H[z_H[n],n-m]x[n].
    \label{sisoPTV}
\end{multline}
An useful alternative representation of this system is given by expressing the input/output signals as vector signals of dimension $K_H$, 
\begin{equation}
    \left\lbrace \begin{array}{l}x_j[r] = x[rK_H+j]\\y_i[k] = y[kK_H+i], \end{array} \right.
\end{equation}
where $i,j\in\{0,1,...,K_H-1\}$. By using the vector-signal representation of the input in Eq.~\ref{sisoPTV} we obtain
\begin{equation}
    y[n] = \sum_{j=0}^{K_H-1}\sum_{r=-\infty}^{\infty} H[z_H[n],n-rK_H-j]x_j[r].
    \label{sisoPTV2}
\end{equation}
Finally, by using the vector-signal representation of the output in Eq.~\ref{sisoPTV2} we have
\begin{equation}
    y_i[k] = \sum_{j=0}^{K_H-1}\sum_{r=-\infty}^{\infty} \bar{H}_{i,j}[k-r]x_j[r],
    \label{MIMO}
\end{equation}
where $\bar{H}$ is a $K_H \times K_H$ matrix given by
\begin{equation}
    \bar{H}_{i,j}[n] = H[i,n K_H+i-j].
\end{equation}
Equation~\ref{MIMO} denotes an interesting equivalence between a SISO PTV and a time-invariant \textit{multiple-input multiple-output} (MIMO) linear system, shown in Fig.~\ref{fig07}(b). Inversely, any MIMO linear system written in the form of Eq.~\ref{MIMO} can be represented as a SISO PTV, by defining the periodic impulse-responses as
\begin{equation}
    H[h,m] = \bar{H}_{h,\mathrm{mod}(m,K_H)}\left[\frac{m-\mathrm{mod}(m,K_H)}{K_H}\right],
    \label{MIMOtoSISO1}
\end{equation}
and the higher-rate input-output signals as
\begin{equation}
    \left\lbrace \begin{array}{l} x[n] = x_{\mathrm{mod}(n,K_H)}\left[(n-\mathrm{mod}(n,K_H))/K_H\right] \\ y[n] = y_{\mathrm{mod}(n,K_H)}\left[(n-\mathrm{mod}(n,K_H))/K_H\right].\end{array} \right.
    \label{MIMOtoSISO2}
\end{equation}

The equivalence shown in Fig.~\ref{fig07} allows the simple calculation of the DTPTV inverse $H^{-1}$, as the inverse of a time-invariant MIMO $K_H \times K_H$ is another MIMO of the same dimension. Basically, the matrix representation of that inverse must satisfy
\begin{equation}
    \sum_{j=0}^{K_H-1}\sum_{m=-\infty}^{\infty} \bar{H}^{(-1)}_{ij}[n]\bar{H}_{jk}[n-m] = \delta_{ik}\delta_{n0},
\end{equation}
where $\delta$ stands for the Kronecker delta. In addition, by using Eqs.~\ref{MIMOtoSISO1} and~\ref{MIMOtoSISO2}, we can write $\bar{H}^{(-1)}$ as a discrete-time SISO PTV. In conclusion, the inverse of a SISO PTV is another SISO PTV of the same period. This conclusion is also valid for continuous-time PTV systems.

\subsection{Square PTV}
\begin{figure}
    \centering
    \includegraphics[scale=0.275]{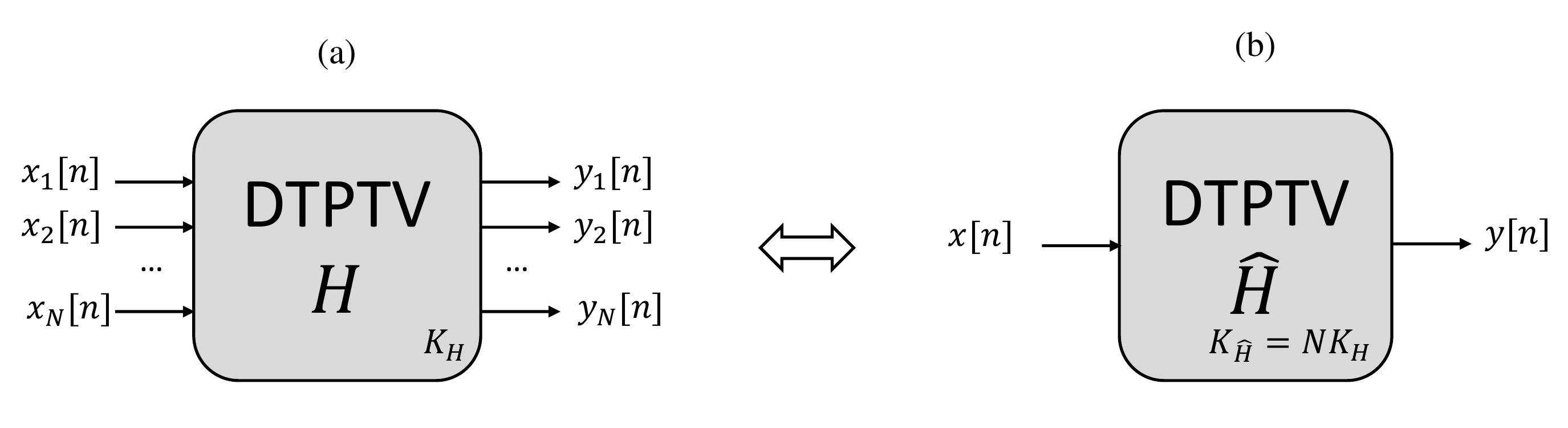}
   \caption{Equivalence between a square PTV and a SISO PTV. a) discrete-time of a square PTV. b) equivalent SISO DTPTV, with $N$ times large discrete period, obtained by considering serializing signals.}
\label{fig08}
\end{figure}

The \textit{square} PTV, shown in Fig.~\ref{fig08}(a), is defined as the PTV system whose number of inputs and number of outputs are equal ($M=N$ in definition of Eq.~\ref{def3}). The discrete-time representation of such system is given by
\begin{equation}
    y_i[n] = \sum_{j=1}^{N} \sum_{m=-\infty}^{\infty} H_{i,j}[\mathrm{mod}(n,K_H),m]x_j[n-m],
    \label{squareDTPTV}
\end{equation}
where $i\in\{1,2,..,N\}$. We define a higher-rate signal to serialize the output of the system, that reads
\begin{equation}
    y[k] = y_i[n], \quad \left\lbrace \begin{array}{l} i = \mathrm{mod}(k,N)+1\\ n = \frac{k-\mathrm{mod}(k,N)}{N}.\end{array}\right.
    \label{serieout}
\end{equation}
In a similar way, we define the serialized input
\begin{equation}
    x[k-r] = x_j[n-m], \quad \left\lbrace \begin{array}{l} j = \mathrm{mod}(k-r,N)+1\\ m = \frac{r-\mathrm{mod}(r,N)}{N}.\end{array}\right.
    \label{serieinput}
\end{equation}
By replacing Eqs.~\ref{serieout} and~\ref{serieinput} into Eq.~\ref{squareDTPTV}, we obtain the TV linear system
\begin{equation}
    y[k] = \sum_{r=-\infty}^{\infty} \hat{H}[k,r] x[k-r],
    \label{aux1}
\end{equation}
where
\begin{equation}
    \hat{H}[k,r] = H_{\mathrm{mod}(k,N)+1,\mathrm{mod}(k-r,N)+1}\left[\mathrm{mod}\left(\frac{k-\mathrm{mod}(k,N)}{N},K_H\right) ,\frac{r-\mathrm{mod}(r,N)}{N} \right].
    \label{Hhat}
\end{equation}
From the definition of Eq.~\ref{Hhat}, it is easy to prove that this system is a DTPTV, since
\begin{equation}
    \hat{H}[k+NK_H,r] = \hat{H}[k,r].
\end{equation}
Consequently, we can rewrite Eq.~\ref{aux1} as
\begin{equation}
    y[k] = \sum_{r=-\infty}^{\infty} \hat{H}[z_{\hat{H}}[n],r] x[k-r],
\end{equation}
where $z_{\hat{H}}[n] = \mathrm{mod}(n,K_{\hat{H}})$, being $K_{\hat{H}} = N K_H$. 

This result means that any square DTPTV can be modeled as a higher-rate SISO DTPTV, as shown in Fig.~\ref{fig08}(b), with a discrete period $N$ times larger. Inversely, we can prove that any SISO DTPTV $\hat{H}$, of period $K_{\hat{H}}$, can be represented as a lower-rate square DTPTV $H$ of period $K_H = K_{\hat{H}}/N$ by defining the parallel inputs/outputs
\begin{equation}
    \left\lbrace \begin{array}{l} x_i[n] = x[nN+i-1]\\ y_i[n] = y[nN+i-1] \end{array} \right.
\end{equation}
and the periodic impulse-responses
\begin{equation}
    H_{i,j}[n,m] = \hat{H}[nN+i-1,mN+j-1].
\end{equation}
Thus, by using the equivalence of Fig.~\ref{fig08}, we can prove that the inverse of a square DTPTV is another DTPTV, analogously to the inverse of a SISO DTPTV. Again, this conclusion is valid for continuous-time systems.

\section{Conclusions}
Starting from a mathematical definition of the periodically time-variant linear systems, we derived simple rules to reduce a circuit, combining different PTVs with time-invariant linear systems, to a single PTV system. By using a frequency-domain analysis of that definition, we obtained a simple formula for the output bandwidth of a PTV, enabling a suitable discrete-time representation of such systems. In addition, we also found interesting equivalences for the DTPTV systems, allowing for the derivation of a meaningful conclusion: the inverse of a square PTV is another square PTV of the same dimension.


\begin{thebibliography}{9}
\bibitem{claasen1982on}
Claasen, T. A. C. M. and W. Mecklenbrauker, \emph{On stationary linear time-varying systems}, IEEE transactions on circuits and systems 29.3:169-184, 1982.

\bibitem{middleton1988adaptive}
Middleton, Richard H. and Graham C. Goodwin, \emph{Adaptive control of time-varying linear systems}, IEEE Transactions on Automatic Control 33.2:150-155, 1988.

\end{thebibliography}
\end{document}